\documentclass[fleqn,12pt,twoside]{article}
\usepackage{espcrc1}

\newcommand{\AmS}{{\protect\the\textfont2
  A\kern-.1667em\lower.5ex\hbox{M}\kern-.125emS}}

\title{
Lowest Order Effective Field Theory for the weak $\Lambda N$
interaction\thanks{Supported by DGICYT BFM2002--01868 (MCyT, Spain)
and SGR2001-64 (Generalitat de Catalunya).}
}

\author{
A. Parre\~no\thanks{Email address: assum@ecm.ub.es}
\address{Dep. ECM, Facultat de F\'{\i}sica,
U Barcelona, E-08028, Barcelona, Spain},
C. Bennhold\address{Center of Nuclear Studies, GWU,
Washington DC, 20052, USA},
B.R. Holstein\address{Dept. of Physics-LGRT, University of Massachusetts,
Amherst, MA 01003, USA}}

\begin{document}
\maketitle
\begin{abstract}
The $|\Delta S|=1$ $\Lambda N$ interaction,
responsible for the decay of hypernuclei,
is studied by means of an Effective Field Theory (EFT) where
the long range physics is described by
pion and kaon exchange mechanisms, and
its short range counterpart is obtained from the most general non-derivative local
four-fermion interaction.
We show that, including the Lowest Order Parity Conserving
(PC) contact terms, allows us to reproduce the total
decay rates for $^5_\Lambda {\rm He}$, $^{11}_\Lambda {\rm B}$ and
$^{12}_\Lambda {\rm C}$
with a reasonable value of ${\hat \chi^2}$, while in order to get a
prediction for the Parity Violating (PV) asymmetry
compatible with experiments, we have to include the Lowest Order PV contact
pieces.
\end{abstract}

\vspace*{0.5cm}
In analogy to the familiar $NN$ phenomenology, the weak $\Lambda N$ interaction
has traditionally been modeled using meson-exchange approaches.
It is well known that the long-range
part of the interaction is well explained by one-pion exchange (OPE),
which can approximately reproduce the hypernuclear non mesonic decay 
rates---$\Gamma$---but not the ratio between the partial rates, 
$\Gamma_n/\Gamma_p =(\Lambda n \to nn)/(\Lambda p \to np)$.
The large energy released in the process ($m_\Lambda - m_N \approx$ 177 MeV)
suggests that short range contributions
could well be important. Along this line, the literature
shows us that the inclusion of the kaon (OKE)
is particularly convenient, due to the
specific interference between OPE and OKE in the PC and PV
channels\cite{PR01,oka2,JOP01}, which substantially enlarges the
$\Gamma_n/\Gamma_p$ ratio, while reducing the total decay rate by a factor of 2.
One way of accounting for short range contributions on top of OKE,
 is to parametrize the physics contained in that region in terms of local
four-fermion interaction terms. This will be the approach followed here, and
we will see that, despite the significant degree
of SU(3) symmetry breaking, we can get an stable chiral expansion
for the $\Lambda N \to NN$ process.
Not considered here is the intermediate-range 2$\pi$-exchange
(two orders higher in the chiral expansion than the corresponding single
pion-exchange piece), nor the last member of the SU(3) Goldstone-boson octet,
the $\eta$, since the strong $\eta NN$ coupling is an order of magnitude
smaller than the corresponding $\pi NN$ and $K \Lambda N$ couplings\cite{etacoup}.

The explicit forms for our OPE and OKE transition potentials can be found
elsewhere\cite{PR01,PRB97}. Here, only the form of the 4-fermion contact
potential, $V_{4P} ({\vec r})$, is presented.
Assume the $\Lambda N$ wave function to be in a $L=0$ relative state. Then,
without any model-dependent assumptions, we can parametrize the two-body
transition through the following Lowest Order (LO) amplitudes:
\footnote{The higher order
$^3S_1 \to ^3D_1$ transition is excluded in the present analysis.}.

\vspace*{0.3cm}
\begin{tabular}{@{}lc}
\hline
$\Lambda N \to NN \;$ {\rm partial \,\, wave}
& {\rm operator} \\
\hline
PC: $\; ^1S_0 \to ^1S_0, \; ^3S_1 \to ^3S_1$  &
${\hat 1} \, \cdot \, \delta^3({\vec r}), \;
{{\vec \sigma_1}\cdot {\vec \sigma_2}} \, \cdot \, \delta^3({\vec r})$ \\
\hline
PV:  $\; ^1S_0 \to ^3P_0, \; ^3S_1 \to ^1P_1$ &
$({\vec \sigma_1} - {\vec \sigma_2}) \cdot \{ \vec{p}_1 - \vec{p}_2 \, , \,
\delta^3({\vec r}) \, \}, \;
({\vec \sigma_1} - {\vec \sigma_2}) \cdot
[ \, \vec{p}_1 - \vec{p}_2\, , \, \delta^3({\vec r})\,  ]$, \\
 & $ {\rm i} \, ({\vec \sigma_1}
\times {\vec \sigma_2}) \cdot
\{ \vec{p}_1 - \vec{p}_2 \, , \, \delta^3({\vec r}) \, \}, \;  {\rm i} \,
({\vec \sigma_1} \times {\vec \sigma_2}) \cdot
[ \, \vec{p}_1 - \vec{p}_2\, , \, \delta^3({\vec r})\, ]$ \\
\hline
PV:  $\; ^3S_1 \to ^3P_1$  &
$({\vec \sigma_1} + {\vec \sigma_2}) \cdot \{ \vec{p}_1 - \vec{p}_2 \, , \,
\delta^3({\vec r}) \, \}, \;
({\vec \sigma_1} + {\vec \sigma_2}) \cdot [ \, \vec{p}_1 - \vec{p}_2\, , \,
\delta^3({\vec r})\,  ]$ \\
\hline
\end{tabular}
\vspace*{0.3cm}

Here, $p_i$ is  the derivative operator acting on the "i{\it th}" particle,
and $\delta^3({\vec r})$ denotes the contact interaction, which we smear by
using a normalized Gaussian form ($f_{ct}(r)$ below), with a typical vector
meson range $\delta \approx 0.36$ fm.
The resulting
leading order $V_{\rm 4P} (\,{\vec r}\,)$ potential for both PV and PC
terms can be written as:

\vspace*{-0.5cm}
\begin{eqnarray}
V_{4P} ({\vec r}) &=&
\left\{ {C_0}^0 + {C_0}^1 \; {\vec \sigma}_1 {\vec \sigma}_2
+ \displaystyle\frac{2 r}{\delta^2}
\left[ {C_1}^0 \; \displaystyle\frac{{\vec \sigma}_1\cdot {\hat r}}{2 {\overline M}}
+ {C_1}^1 \; \displaystyle\frac{{\vec \sigma}_2\cdot {\hat r}}{2 M}
+ {C_1}^2 \;
\displaystyle\frac{({\vec \sigma}_1 \times {\vec \sigma}_2)
\cdot {\hat r}}{2 {\tilde{M}}} \right] \right\} \nonumber \\
&\times& f_{ct} (r)
\; \times \left[
C_{\rm IS} \, {\hat 1} + C_{\rm IV} \,
{\vec \tau_1}\cdot {\vec \tau}_2
\right] \, {\rm ,}
\label{eq:4ppotr}
\end{eqnarray}
where $\overline{M} = (M_N+M_\Lambda)/2$,
$\tilde{M}= (3 M+M_\Lambda)/4$,
${C^j}_i$ is the j$th$ LEC at i$th$ order, and
the last factor represents the
isospin part of the 4-fermion interaction\footnote{Note that
we only allow for $\Delta I=1/2$ transitions.}, containing
isoscalar ($C_{\rm IS}$) and isovector ($C_{\rm IV}$) pieces.

Standard nuclear structure techniques\cite{PRB97} are used to
decouple from our initial hypernuclei a $\Lambda N$ pair, which
will connect through the two-body potential to a $NN$ final pair.
In addition, the effects of the strong interaction are accounted for
by using the realistic (one-boson-exchange) NSC97f baryon-baryon
interaction model\cite{nij99}.

Our results are shown in Table~\ref{OBS}.
It is noticeable that including the LO PV terms, which are one order
higher in the chiral expansion,
does not substantially alter the previously fitted LO PC coefficients, thus
supporting the validity of our expansion.
Moreover, they barely modify the total and partial
rates but significantly affect the asymmetry, ${\cal A}$,
as one should expect for an
observable defined by the interference between PV and PC amplitudes.
Finally, we find coefficients of natural size with significant error bars,
reflecting the level of experimental uncertainty.

The present study then supports the validity of the EFT framework
for nonmesonic hypernuclear weak decay, and we hope to be able to
much better constrain the parameters in such studies
with the
next generation of high-precision experimental data,
currently under analysis.

\begin{table}[hbt]
\caption{
Results obtained for the weak decay observables and LEC,
when a fit to the $\Gamma$ and $n/p$ for
$^5_\Lambda {\rm He}$, $^{11}_\Lambda {\rm B}$ and
$^{12}_\Lambda {\rm C}$ is performed.
The values between parenthesis have been obtained including
the helium asymmetry, ${\cal A} (^5_\Lambda {\rm He})$, in the fit.}
\label{OBS}
\renewcommand{\tabcolsep}{0.8pc} % enlarge column spacing
\renewcommand{\arraystretch}{1.2} % enlarge line spacing
\begin{tabular}{@{}lccccc}
\hline
 & $\pi$ & $+ K$ & $+$ LO PC & $+$ LO PV& EXP: \\
\hline
$\Gamma (^5_\Lambda {\rm He})$ & $0.42$ & $0.23$
& $0.43$ & $0.44$ ($0.44$)
 & $0.41 \pm 0.14$\cite{Szy91} \\
 & & & & & $0.50 \pm 0.07$\cite{No95} \\
$\Gamma_n/\Gamma_p (^5_\Lambda {\rm He})$ & $0.09$ & $0.50$
& $0.56$ & $0.55$ ($0.55$)
& $0.93\pm 0.55$\cite{Szy91} \\
& & & & & $0.50 \pm 0.10$\cite{kek02} \\
${\cal A} (^5_\Lambda {\rm He})$ & $-0.25$ & $-0.60$
& $-0.80$ & $0.15$ ($0.24$)
 & $0.24 \pm 0.22$\cite{Ajim00} \\
\hline
$\Gamma (^{11}_\Lambda {\rm B})$ & $0.62$ & $0.36$
& $0.87$ & $0.88$ ($0.88$)
 &  $0.95  \pm 0.14$\cite{No95} \\
$\Gamma_n/\Gamma_p (^{11}_\Lambda {\rm B})$ & $0.10$ & $0.43$
 & $0.84$ & $0.92$ ($0.92$)
 & $1.04^{+0.59}_{-0.48}$\cite{Szy91} \\
${\cal A} (^{11}_\Lambda {\rm B})$ & $-0.09$ & $-0.22$
& $-0.22$ & $0.06$ ($0.09$)
 & $-0.20 \pm 0.10$\cite{Aj92} \\
\hline
$\Gamma (^{12}_\Lambda {\rm C})$ & $0.74$ & $0.41$
& $0.95$ & $0.93$ ($0.93$)
 & $1.14\pm 0.2$\cite{Szy91} \\
 & & & & & $0.89 \pm 0.15$\cite{No95} \\
 & & & & & $0.83\pm 0.11$\cite{Bhang} \\
$\Gamma_n/\Gamma_p (^{12}_\Lambda {\rm C})$ & $0.08$ & $0.35$
& $0.67$ & $0.77$ ($0.77$)
 & $0.87 \pm 0.23$\cite{Ha02} \\
${\cal A} (^{12}_\Lambda {\rm C})$ & $-0.03$ & $-0.06$
& $-0.05$ & $0.02$ ($0.03$)
 & $-0.01 \pm 0.10$\cite{Aj92} \\
\hline
${\hat \chi}^2$ & & & $0.98$ & $1.50$ ($1.16$) & \\
\hline
$C_0^0$ & & &  $-1.51 \pm 0.38$ & $-1.09 \pm 0.36$
($-1.02 \pm 0.35$) & \\
$C_0^1$ &  & & $-0.86 \pm 0.24$ & $-0.63 \pm 0.35$
($-0.57 \pm 0.29$) & \\
$C_1^0$ &  & & $---$ & $-0.45 \pm 0.42$
($-0.47 \pm 0.17$) & \\
$C_1^1$ &  & & $---$  & $0.17 \pm 0.22$
($0.20 \pm 0.19$) & \\
$C_1^2$ &  & & $---$ & $-0.48 \pm 0.20$
($-0.48 \pm 0.22$) & \\
$C_{IS}$ &  & & $5.08 \pm 1.27$ & $5.69 \pm 0.74$
($5.83 \pm 0.82$) & \\
$C_{IV}$ &  & & $1.47 \pm 0.39$ & $1.49 \pm 0.23$
($1.52 \pm 0.24$) & \\
\hline
\vspace*{-1.0cm}
\end{tabular}
\end{table}


\begin{thebibliography}{9}

\bibitem{PR01}
A. Parre\~no and A. Ramos, Phys. Rev. C {\bf 65}, 015204 (2002).

\bibitem{oka2} K. Sasaki, T. Inoue, and M. Oka, Nucl. Phys.
{\bf A678}, 455 (2000).

\bibitem{JOP01} E. Jido, E. Oset and J.A. Palomar,
Nucl. Phys. {\bf A694}, 525-555 (2001).

\bibitem{etacoup}
G. Penner and U. Mosel,{\em Phys. Rev.} {\bf C 66} (2002) 055211;
L.Tiator, C. Bennhold, and S.S. Kamalov, Nucl. Phys. {\bf A580}, 455
(1994).
%; Shi-Lin Zhu, {\em Phys. Rev.} {\bf C 61} (2000) 065205.

\bibitem{PRB97}
A. Parre\~no, A. Ramos, and C. Bennhold, Phys. Rev.
C {\bf 56}, 339 (1997).

\bibitem{nij99}
V.G.J. Stoks and Th.A. Rijken, Phys. Rev. C {\bf 59}, 3009 (1999);
Th.A. Rijken, V.G.J. Stoks and Y. Yamamoto, Phys. Rev. C {\bf 59}, 21-40 (1999).

\bibitem{Szy91}
J.J. Szymanski et al., Phys. Rev. C {\bf 43}, 849 (1991).

\bibitem{No95}
H. Noumi et al., Phys. Rev. C {\bf 52}, 2936 (1995).

\bibitem{kek02} H. Outa {\sl et al.}, Proposal of KEK--PS E462 (2000);
H. Outa, talk presented at the {\em XVI PaNic02 Int.
Conference}, Osaka (Japan), Sept 30 -- Oct 4, 2002.

\bibitem{Ajim00}
S. Ajimura et al., Nucl. Phys. {\bf A663}, 481 (2000).

\bibitem{Aj92}
S. Ajimura et al., Phys. Lett. {\bf B282}, 293 (1992).

\bibitem{Bhang}
H. Bhang {\em et. al.}, Phys. Rev. Lett. {\bf 81}, 4321 (1998).

\bibitem{Ha02}
Erratum of Ref.:
O. Hashimoto {\sl et al.}, {\em Phys. Rev. Lett.} {\bf 88},
042503 (2002).

\end{thebibliography}
\end{document}